\documentclass[aps,pre,superscriptaddress,amsmath,amssymb,floatfix]{revtex4} 

\usepackage{graphics}
\usepackage{amsmath}
\usepackage{amssymb}

\newcommand{\dt}{\Delta t}

\begin{document}
\title{Effective interactions and large deviations in stochastic processes}
\author{Robert L. Jack}
\affiliation{Department of Physics, University of Bath, Bath BA2 7AY, United Kingdom}
\author{Peter Sollich}
\affiliation{Department of Mathematics, King's College London, Strand, London WC2R 2LS, United Kingdom}

\begin{abstract}
We discuss the relationships between large deviations in stochastic systems,
and ``effective interactions'' that induce particular rare events.
We focus on the nature of these effective interactions in physical systems with many
interacting degrees of freedom, which we illustrate by reviewing several recent studies.  We describe the 
connections between effective interactions, large deviations at ``level 2.5'', and the theory of optimal control.  Finally,
we discuss possible physical applications of variational results associated with those theories. 
\end{abstract}

\maketitle

\newcommand{\ee}{\mathrm{e}}
\newcommand{\CC}{\mathcal{C}}
\newcommand{\WW}{\mathbb{W}}
\newcommand{\HH}{\mathbb{H}}
\newcommand{\tobs}{t_\mathrm{obs}}

\section{Introduction}

Rare events are important in many physical settings: classic examples include phase transformation, 
protein-folding, and chemical reactions~\cite{auer01,sear07,ren05,hanggi90}.   In those cases, a system makes a transition between two 
distinct states, and a variety of analytical and computational tools are available~\cite{hanggi90,wales-book,e05,tps,allen06}.
Here, we focus on a different class of rare events, where systems behave in an unusual fashion over
an extended period of time.  Specifically, we consider the probability of trajectories in which time-averaged
quantities remain far om their typical (equilibrium) values.  If the system is ergodic, the probabilities of such events decay to zero as the length
of trajectory goes to infinity: the rate of this decay is described by the  mathematical theory of large deviations~\cite{touchette09}.  Recent studies of these large deviations have
provided insights into fluctuation theorems~\cite{galla95,lebowitz99}, glassy systems~\cite{garrahan07,garrahan09,hedges09,speck12}, 
protein-folding~\cite{weber13,mey14,weber14}, chaotic dynamical systems~\cite{tailleur07,lam09} and interacting particle 
models~\cite{derrida98,bertini01,bodineau04,bertini14}.

It turns out that the rare trajectories of interest in these systems
can be characterised as typical trajectories for a certain modified system~\cite{dv75,evans04,evans05,maes08,js10,chetrite14-hp}, 
which we refer to here as the ``auxiliary model''.  
The auxiliary model inherits many of its important properties from the original system of interest: if the original model has the Markov
property then so does the auxiliary model.  In many cases, the auxiliary model inherits the symmetries of the original model, and other properties
like kinetic constraints are also preserved~\cite{js10}.  

The existence of this auxiliary model raises important questions for characterisation of rare events.  In particular, it means that by adding
a particular set of interactions to the original model, one may drive the system to realise these rare events.  In fact, these interactions
can be shown to be the ``optimal'' ones for realising the rare events of interest, in a certain precise sense~\cite{fleming92,fleming05,hartmann12,kappen13} 
(see Section~\ref{sec:opt}, below).   It is therefore
of great interest to characterise these interactions. For example, in glassy systems, they can stabilise ``amorphous solid
states''~\cite{hedges09,speck12} that are otherwise only metastable -- the nature of the interactions required to achieve this 
is a long-standing question in the field.  
In protein-folding systems, effective
interactions might stabilise the native state, or they might favour misfolded states~\cite{mey14,weber14}: 
understanding how these states can be characterised (and suppressed) is of vital importance in that context.  

In this paper, we survey some key results that are related to the existence and nature of these auxiliary models, and the effective interactions that they encode.  Our aim is to draw together ideas from several different contexts and to give a (non-rigorous) presentation that highlights the central outstanding questions, and possible routes to solving them.  In Section~\ref{sec:setting}, we describe the setting for our main results.  Section~\ref{sec:illustrate} illustrates the kinds of phenomena that we are interested in, through a summary of some recent numerical results.  Then, in Section~\ref{sec:noneq}, we describe some theoretical results, including the relationship to large deviations at ``level-2.5''~\cite{chetrite14-l2p5} and to optimal control theory~\cite{fleming92,fleming05,hartmann12,kappen13}.  These results have not yet been exploited very far in the physics context -- we highlight possibilities for future progress along these directions.  Section~\ref{sec:conc} gives a brief summary and outlook.

\section{Basic theory}
\label{sec:setting}

In this section, we collect some key results related to large deviations in stochastic processes.  Many of these results have
been derived independently in different contexts and by different groups. Here we follow the presentation of~\cite{lecomte07jsp,garrahan09,js10};
further details and references can be found in those works.

\subsection{Models and master equations}

We consider a Markov process in continuous time, on a (finite) discrete state space with configurations $\{ \CC \}$.
For example, one can consider a lattice of Ising spins, or a simple particle model such as the asymmetric exclusion process.
In practical settings, one is often interested in the thermodynamic limit, where the size of the state space is taken to infinity, for example by considering spins on lattices of increasing size.  Alternatively, one may consider diffusive processes described by Langevin equations
(stochastic differential equations).  These may typically be obtained from lattice models by a continuum limit: one defines a process on a discrete lattice and then takes the lattice spacing to zero, rescaling time in an appropriate way to ensure diffusive behaviour.  Our restriction to finite state spaces means that the following
analysis may not always be valid on taking thermodynamic or continuum limits -- in typical cases we expect our results to remain valid in such limits,
but this is not guaranteed.  

The transition rates between configurations of the system are
$W(\CC'\leftarrow \CC)$. Let $P(\CC,t)$ be the probability that the system is in configuration $\CC$ at time $t$: this quantity evolves by a Master equation
\begin{equation}
\partial_t P(\CC,t) = - r(\CC) P(\CC,t) + \sum_{\CC'} W(\CC\leftarrow \CC') P(\CC',t) 
\label{equ:master}
\end{equation}
where $r(\CC) = \sum_{\CC'} W(\CC'\leftarrow \CC)$ is the ``escape rate'' from configuration $\CC$.
We assume that the process is irreducible, which ensures ergodicity, since the state space is finite.
It is also useful to identify the subclass of these models that obey ``detailed balance''.  For these models, there exists a ``potential''
$E_\CC$ such that 
\begin{equation}
W(\CC\leftarrow \CC')\ee^{-E_{\CC'}} = W(\CC'\leftarrow \CC)\ee^{-E_\CC},
\label{equ:detbal}
\end{equation}  
for all $\CC$ and $\CC'$.  
Models with this property   
 have time-reversal symmetric (``equilibrium'') steady states, in which the probability
distribution over configurations is $p(\CC) \propto \ee^{-E_\CC}$.  

\subsection{Large deviations and biased ensembles}

The rare events that we consider are defined by the choice of an observable, which may be one of two types.
A trajectory of the system consists of the (ordered) set of states which the system visits, and the times at which transitions (jumps) between states take
place.
The first type of observable takes the general form 
\begin{equation}A = \sum_{{\rm jumps}\, \CC\to\CC'} \alpha(\CC'\leftarrow\CC) \end{equation}
where the sum runs over all transitions 
within the trajectory and the $\alpha(\CC'\leftarrow\CC)$ are a given set of numbers. For example, if $\alpha=1$ for all pairs of configurations then $A$ is the total
number of configuration changes in the trajectory.  
 The second type of observable is the time integral of a state-dependent quantity
\begin{equation} 
B=\int_0^{\tobs}\mathrm{d}t\, b(\CC(t)). 
\end{equation}
where $\CC(t)$ is the configuration of the system at time $t$.
For large $\tobs$, the probability distribution
of $B$ generically has a large deviation form:
\begin{equation}
p(B) \sim \exp[-\tobs \phi(B/\tobs)]
\label{equ:b-phi}
\end{equation}
where $\phi(b)$ is known as a rate function.  
 A similar expression holds for the distribution of $A$.  
[The precise meaning of (\ref{equ:b-phi}) is that $\lim_{\tobs\to\infty} \tobs^{-1} \ln p(B=b\tobs) = \phi(b)$;
the `$\sim$' symbol is used in this sense throughout this article.]
 The main question of interest in the following is:
what kinds of dynamical trajectory dominate the distribution $p(B)$ when $B$ is not equal to its typical (steady-state) value?

To obtain information about these trajectories, it is convenient to write a biased probability
distribution over the possible trajectories of the model:
\begin{equation}
\mathrm{Prob}[\CC(t);s] = \mathrm{Prob}[\CC(t);0] \cdot \frac{\ee^{-sB[\CC(t)]}}{Z(s,\tobs)}
\label{equ:s-ens}
\end{equation}
where $\mathrm{Prob}[\CC(t);0]$ is the unbiased (steady-state) probability distribution over trajectories $\CC(t)$, the parameter $s$ sets the strength of the bias, and
$Z=\langle \ee^{-sB} \rangle_0$ resembles a partition function.
Hence, the average any observable $O$ within this generalised ensemble is 
\begin{equation}\langle O \rangle_s = \frac{ \langle O \ee^{-sB}\rangle_0 }{ Z(s,\tobs) }. \end{equation}
It may be shown~\cite{chetrite13} that averages within this biased ensemble are the same as those in an ensemble
in which the value of $A$ (or $B$) is constrained to a particular
value. 
[Note that this equivalence is assured only in systems with finite state
spaces, in which case the free energy $\psi(s)$ is analytic and convex. In systems with infinite state spaces, dynamical phase transitions~\cite{bodineau05,lecomte07jsp,garrahan07} may mean that trajectories which are representative of some values of $A$ (or $B$) cannot be obtained within biased ensembles of the form given in (\ref{equ:s-ens}).] 

To analyse these biased ensembles, one considers the probability that a system is in configuration $\CC$ at time $t$,
and that the observable $B$ has a particular value associated with the trajectory up to time $t$~\cite{lebowitz99,lecomte07jsp,garrahan09}.  
(The analysis for observables of type $A$ is similar.)
If this probability is $p(\CC,B,t)$
then we define $p(\CC,s,t) = \int\mathrm{d}B \, p(\CC,B,t) \ee^{-sB}$. This quantity evolves by an equation which is formed from 
(\ref{equ:master}) by replacing $P(\CC,t)$
with $P(\CC,s,t)$, and adding a term
$-s b(\CC) p(\CC,s,t)$ to the right hand side. 
The resulting equation is linear in $P$ so it is useful to write it formally as
\begin{equation}
\partial_t |P\rangle = \WW(s) |P\rangle
\label{equ:wsp}
\end{equation}
where $\WW(s)$ is an operator (matrix) with  
diagonal elements $-r(\CC) - sb(\CC)$ and
off-diagonal elements $W(\CC'\leftarrow\CC)$.  In the case of type-$A$ observables, the parameters $\alpha(\CC'\leftarrow \CC)$ appear in the
off-diagonal elements via multiplicative factors $\ee^{-s\alpha}$~\cite{lecomte07jsp,garrahan09}.  Note that (\ref{equ:wsp}) resembles a master equation, but it does not 
conserve probability (in the sense that $\sum_\CC p(\CC,s,t)$ is not constant under the time evolution).

\subsection{Connection between type-$A$ and type-$B$ observables}

We note at this point that the operator $\WW(s)$ fully specifies the probability distribution in (\ref{equ:s-ens}), up to
possible boundary terms that we will neglect in the following (see also Sec.~\ref{sec:aux}, below).
This means that if two processes are associated with the same operator $\WW(s)$, then they have the same behaviour.
It follows that ensembles defined by type-$A$ observables can be given alternative definitions in terms of type-$B$ observables,
but for a different underlying stochastic model.

For example, suppose that a model has transition rates $W(\CC'\leftarrow \CC)$ and is biased by an observable of type $A$.
Then, the same operator $\WW(s)$ can be obtained by considering a different model with transition rates
$\tilde{W}(\CC'\leftarrow\CC) = W(\CC'\leftarrow \CC)\ee^{-s\alpha(\CC'\leftarrow \CC)}$, biased by an observable
$\tilde{B} = s^{-1} \int\mathrm{d}t [ r(\CC_t) - \tilde{r}(\CC_t)]$ where $\tilde{r}(\CC) = \sum_{\CC'} \tilde{W}(\CC'\leftarrow\CC)$:
see for example~\cite[Appendix B]{garrahan09}.
A similar transformation means that any $B$-biased process can always
be re-written as an $A$-biased one. 
(This requires that $r(\CC)+sb(\CC)>0$ for all configurations, which in finite state spaces can always be achieved by including an appropriate constant shift in $b(\CC)$.)
Hence, in the following, we sometimes state results either for type-$A$ or type-$B$ observables, since the results 
for the other type can always be derived by an appropriate transformation.

\subsection{Auxiliary models}
\label{sec:aux}

Given a model [specified by rates $W(\CC'\leftarrow\CC)$] and an observable [specified by the  $\alpha(\CC'\leftarrow\CC)$ or $b(\CC)$], one may
always define an auxiliary model whose steady state distribution of trajectories is close to (\ref{equ:s-ens}).  [A precise characterisation of this
``closeness''
is given in (\ref{equ:ps-aux}) below].
For observables of type $B$, the transition rates of the auxiliary model are~\cite{evans04,evans05,js10,chetrite14-hp}
\begin{equation}
W^\mathrm{aux}(\CC'\leftarrow\CC) = u_{\CC'} W(\CC'\leftarrow\CC) u_{\CC}^{-1} 
\label{equ:waux}
\end{equation}
where the $u_\CC$ are obtained by solving an eigenvalue equation for the operator $\WW(s)$. Specifically, $\langle u|$ is the left eigenvector associated with the smallest eigenvalue of $-\WW(s)$: 
\begin{equation}
 \langle u |(-\WW(s)) = \psi(s) \langle u| .
\label{equ:eval}
\end{equation}
Here $\psi(s)$ is a dynamical free energy, related to the dynamical partition function by $Z(s,\tobs) \sim \ee^{-\tobs \psi(s)}$.
We note the connection of (\ref{equ:waux}) to Doob's $h$-transform~\cite{stroock}, which is one of the earliest results connecting rare events to auxiliary models of
this kind.  Similar results also appear in other kinds of biased rare-event problems~\cite{maes08,hartmann12,cameron14}. 

Eq.~(\ref{equ:waux}) motivates us to define an ``effective potential''
\begin{equation}
\Delta V_\CC = -2 \ln u_\CC  .
\label{equ:eff-pot}
\end{equation}
With this definition, $W^\mathrm{aux}(\CC'\leftarrow\CC) = W(\CC'\leftarrow\CC) \ee^{(\Delta V_{\CC} - \Delta V_{\CC'})/2}$, which can be interpreted as
a modification of the original transition rates according to the change of the effective potential in a transition.
For type-$A$ observables, the analogue of (\ref{equ:waux}) is
\begin{equation}
W^\mathrm{aux}(\CC'\leftarrow\CC) = u_{\CC'} W(\CC'\leftarrow\CC) \ee^{-s\alpha(\CC'\leftarrow\CC)} u_{\CC}^{-1}
\label{equ:waux-a}
\end{equation}

To see the relation between $W^{\rm aux}$ and $\WW(s)$, we define an operator $\WW^{\rm aux}$ whose
off-diagonal elements are the $W^\mathrm{aux}(\CC'\leftarrow\CC)$ and whose diagonal elements
are $-r^{\rm aux}(\CC)$, with escape rates $r^{\rm aux}(\CC) = \sum_{\CC'} W^\mathrm{aux}(\CC'\leftarrow\CC)$.  If we also define
$\hat{u}$ to be a diagonal
operator whose elements are the $u_\CC$, it follows~\cite{js10} that
\begin{equation}
\WW^{\rm aux} = \hat{u} \WW(s) \hat{u}^{-1} + \psi,
\label{equ:waux-op}
\end{equation}
which
holds for both type-$A$ and type-$B$ observables. 

With these definitions, the trajectory measure for the steady state of the auxiliary model, $\mathrm{Prob}[\CC(t)];{\rm aux}]$, is related
to the biased ensemble (\ref{equ:s-ens}) as
\begin{equation}
\mathrm{Prob}[\CC(t);s] = \mathrm{Prob}[\CC(t);{\rm aux}] \cdot \frac{ \ee^{[\Delta V_{\CC(\tobs)} - \Delta V_{\CC(0)}]/2} }{ Z^{\rm aux} } \cdot \frac{ p_0(\CC(0)) }{ p^{\rm aux}(\CC(0)) }
\label{equ:ps-aux}
\end{equation}
where $Z^{\rm aux}$ is a normalisation constant, and the final factor on the rhs 
is the ratio of steady state probabilities of the initial configuration $\CC(0)$, in the original
process [$p_0(\CC(0))$] and the auxiliary process  [$p^{\rm aux}(\CC(0))$].
Eq.~(\ref{equ:ps-aux}) is most easily derived via direct construction of the various $\mathrm{Prob}[\CC(t)]$.
For example, if the biasing obervable is of type $A$ then we have
\begin{equation}\mathrm{Prob}[\CC(t),s]
=\left[\prod_{k=1}^K  \ee^{-(t_k-t_{k-1})r(\CC_{k-1})} \ee^{-s\alpha(\CC_k\leftarrow \CC_{k-1})} W(\CC_k\leftarrow \CC_{k-1})\right]
\ee^{-(\tobs-t_{K})r(\CC_{K})} p_{\rm 0}(\CC_0)  \frac{1}{Z(s,\tobs)}
\label{equ:ptraj}
\end{equation} 
where
the trajectory is composed of configurations $\CC_0,\CC_1,\dots,\CC_K$, with configuration changes at times $t_1,t_2,\dots t_K$,
we define $t_0=0$, and $p_0(\CC)$ is the probability of finding configuration $\CC$ in the steady state of the original (unbiased) model.  
A similar construction of the analogous probability density for the auxiliary process then yields (\ref{equ:ps-aux}).  For a detailed
analysis, see~\cite{chetrite14-hp}.

Since all differences between the probability distributions in (\ref{equ:ps-aux}) come from the initial and final
states, we expect that for long trajectories, the distributions 
${\rm Prob}[\CC(t);s]$ and ${\rm Prob}[\CC(t);{\rm aux}]$ will differ only through initial and final ``transient'' regimes, and that their behaviour will
be the same in the intermediate-time regime for which $t\gg 1$ and $\tobs-t\gg 1$.  The transient regimes are discussed in more
detail in~\cite{garrahan09} and also in~\cite{chetrite14-hp}, where it was shown how a set of time-dependent auxiliary rates can lead
to exact correspondence between the auxiliary and biased processes.

It is useful to note that (for type-$B$ observables) 
\begin{equation}
u(\CC) \propto \lim_{\tobs\to\infty} \langle \ee^{-sB + \tobs\psi(s)} \rangle_{\CC,0} 
\end{equation}
where the average is taken with respect to the unbiased dynamics, for a system initialised in configuration $\CC$~\cite{evans04,evans05,js10,nemoto14}. The term $\tobs\psi(s)$ in the exponent ensures that the average does not grow or decay exponentially in time, because from the definition of $Z$ and its link to the dynamical free energy one has
\begin{equation}
\ee^{-\tobs \psi(s)} \sim Z(s,\tobs) = \langle \ee^{-sB} \rangle_0  .
\label{equ:z-psi}
\end{equation}

\subsection{Biased ensembles with time-reversal symmetry}

In cases where the biased ensembles are symmetric under time-reversal, the eigenvalue problem (\ref{equ:eval}) may be simplified:
it reduces to finding
the largest eigenvalue of a symmetric matrix.  The most common situation in which this occurs is when the unbiased model obeys detailed balance,
and the biasing observable is either of type-$B$, or of type-$A$ with $\alpha(\CC'\leftarrow\CC) = \alpha(\CC\leftarrow\CC')$ for 
all $\CC$ and $\CC'$.  In this case one has simply~\cite{garrahan09,js10} 
\begin{equation}
\psi = \min_{|x\rangle} \frac{ \langle x | \ee^{\hat{E}/2} (-\WW(s)) \ee^{-\hat{E}/2} | x\rangle}{ \langle x | x \rangle }
\label{equ:var-H}
\end{equation}
where $\hat{E}$ is a diagonal operator whose elements are the energies $E_\CC$ that appear in the detailed balance relation (\ref{equ:detbal}),
and the maximisation is over vectors with elements $x_\CC$.  The maximum occurs when $x_\CC = u_\CC\ee^{-E_\CC/2}$
so this variational
result allows direct estimation of the effective interactions.  Generalisations of this result to cases without time-reversal symmetry will
be discussed in Section~\ref{sec:noneq} below.

\section{Illustrative results from model systems}
\label{sec:illustrate}

Having introduced the general features of biased ensembles of trajectories, 
we now return to our original focus on complex systems with many interacting degrees of freedom.
In these cases, it is not usually possible to solve the eigenproblem (\ref{equ:eval}) in order to obtain the $u_\CC$.  
Further, even if
this eigenvector could be obtained exactly, it typically has such a large dimensionality that it does not provide
direct information about the physical nature of effective interactions in the system.  To illustrate these physical ideas, we now recall
some recent results on the physical features of effective interactions in biased ensembles, for different model systems.

\subsection{Glass-forming systems}
\label{sec:kcm}

Kinetically constrained models consist of interacting spins (or particles) in which local rules mean that only a subset of spins are able to flip at any 
given time step~\cite{rs03,gst11}
These models
provide
simple descriptions of glass-forming liquids~\cite{gc10}.  The ``mobile''
subset of spins changes with time, and the system is ergodic on long time scales.  The dynamical motion in these systems can be complex and co-operative,
even if their static (thermodynamic) properties are very simple.  

In these systems, it is typically possible to construct configurations in which
the subset of mobile spins remains finite in the limit of large system size.  In this case, if one considers the large deviations of
 the total number of spin flips in a trajectory (type-$A$ observable with all $\alpha=1)$, it can be shown from (\ref{equ:var-H}) that (i)
$\lim_{N\to\infty}\psi/N \leq 0$ where $N$ is the system size, (ii) this bound is saturated for all $s>0$, and (iii) the effective interaction in this case drives the system
into configurations with a finite number of mobile spins.
It follows that these systems have dynamical phase transitions at $s=0$~\cite{garrahan07,garrahan09}.  The dominant
feature of the effective interactions for $s>0$ is a very strong suppression of mobile spins, although the detailed nature of the effective
interactions that produce this suppression is not known.

Similar phase transitions exist  
in fully-connected (``mean-field'') spin-glass models with large numbers of metastable states~\cite{jack10}, and there is also numerical evidence
for them in atomistic models of glass-forming liquids~\cite{hedges09,speck12},  but the nature of the effective interactions again remains unclear.

A recent study of a particular kinetically
constrained model (the East model~\cite{rs03}) highlights the complex effective interactions that can appear even in simple systems.  
On biasing this model to low activity, one observes the dynamical phase transition discussed above.
However, if one biases instead to high activity, one observes a hierarchy of responses that mirror the ``aging'' behaviour of the same model~\cite{jack14}. 
(Aging behaviour occurs
when the system is initialised at high temperature followed by dynamical relaxation at low temperature.)  The dominant features of these states are 
(i) effective interactions that are long-ranged even for weak biases $s$, and (ii) a hierarchy of length scales associated
with different relaxation processes within the system.

\subsection{Exclusion processes}

There have been many studies of large deviations in exclusion processes, in which particles move on a lattice, with at most one particle per site. 
Effective interactions in biased ensembles have been considered in relatively few cases; two examples are  
the limits of maximal dynamical activity or maximal current, where the effective interactions can be found exactly~\cite{popkov11}.  These interactions are dominated
by a long-ranged repulsion between particles: the system can be mapped to a ``one-component plasma'' of positively-changed 
particles interacting by Coulomb-like forces.  The result of these long-ranged forces is that the system becomes ``hyperuniform''~\cite{jack14-hyper} 
-- density
fluctuations on large length scales are strongly suppressed~\cite{torquato03}.  Such correlations occur in a variety of non-equilbrium systems~\cite{gabrielli2003,zachary,chicken2014}, but they
are forbidden in equilibrium systems with short-ranged forces.  

Biasing exclusion processes to small activity can also result in phase transitions
into inhomogeneous states~\cite{bodineau04,bertini05,bodineau05}, although the effective interactions associated with these states have not been investigated in detail.
Similar behavior can occur in simple models of heat conduction~\cite{hurtado14}.

\subsection{Numerical results}

As well as these analytic results, there are several numerical methods that allow large deviations to be investigated.  Briefly, \emph{transition path 
sampling}~\cite{tps} is a computational method for sampling trajectories of systems according to general path ensembles, 
including examples such as (\ref{equ:s-ens})~\cite{merolle05,hedges09}.
The method is most easily implemented for processes obeying detailed balance, although generalisations are possible~\cite{crooks01}.  Alternatively
the \emph{cloning} method was developed specifically to study large deviations~\cite{giardina06,tailleur07,lecomte07jsm} and is not restricted
to systems with time-reversal symmetry -- it involves many copies (``clones'') of the system evolving in parallel.
Finally, a third method was proposed recently by Nemoto and Sasa~\cite{nemoto14}, which involves direct estimation of the auxiliary rates in (\ref{equ:waux-a}), in a manner reminiscent
of thermodynamic integration.

These methods have provided a number of interesting insights, especially for models that are not tractable analytically.
Examples include model protein-folding systems~\cite{mey14}, where biased ensembles are dominated by ``misfolded'' states, 
reminiscent of the low-activity states discussed in Sec.~\ref{sec:kcm}.
Similar results can also be obtained in protein systems for which Markovian effective descriptions are available -- if the resulting state
space is sufficiently small then large deviations can be analysed by exact diagonalisation of the operator $\WW(s)$~\cite{weber13}.  One again
finds that the effective interactions stabilise misfolded metastable states~\cite{weber14}. 

Numerical
methods have also been used to study the competition between chaotic and periodic behaviour in dynamical systems~\cite{tailleur07,lam09}.  In particular,
even if a system's steady state is chaotic, its large deviations may be characterised by periodic trajectories, which allow the system
to avoid ``equilibration'' into an ergodic state.

We emphasise that the path sampling and cloning methods do not provide direct information about effective interactions, 
and even the method of~\cite{nemoto14} typically requires an
approximate parameterisation of these interactions to be chosen before starting the analysis.  However, the methods do yield representative
configurations of the biased system, which at least provide qualitative insights into the underlying interactions.  We believe that further development of
methods in this area is a useful area for further study.

\subsection{General principles}

We identify two general principles from the illustrative examples above.  Firstly, biased ensembles of trajectories
often contain correlations that are very unusual in equilibrium systems.  The hyperuniform states found in exclusion processes are stabilised
by long-ranged effective interactions~\cite{popkov11,jack14-hyper} -- these might not have been anticipated given the simple local rules and the simple bias to high activity.
Similarly, the long-ranged correlated states found in the East model biased to high activity do not at all resemble the equilibrium
state of that system~\cite{jack14}, and nor do the periodic (non-chaotic)
trajectories found in some dynamical systems~\cite{tailleur07,lam09}.  We emphasise that biased states are optimised with respect to global observables ($A$ or $B$)
that depend on the whole system, integrated over a long period of time, so there is no general reason to expect effective interactions
to be the short-ranged forces that are familiar from equilibrium settings.   So one may expect to find new and unusual phenomena on
investigating large deviations.

Secondly, effective interactions are often linked with underlying metastable states
in a system -- biasing to low activity often drives the system into ``glassy'' metastable states, as found in kinetically-constrained models~\cite{garrahan07,garrahan09},
atomistic glass-formers~\cite{hedges09,speck12}, and proteins~\cite{weber13,mey14,weber14}.  Given the variational principle (\ref{equ:var-H}), this may not be suprising -- the low-lying eigenvalues
of the operator $-\WW(0)$ are naturally linked with metastable states and phase transitions, so weak perturbations can be expected
to lead to hybridisation of these states with the dominant eigenvector.  However, the use of large deviation
methods to further analyse dynamical metastability and glassy behaviour seems promising.  For example, recent work on biased ensembles
in quantum systems also highlights the importance of quiescent (inactive) states that couple weakly to their enviroment~\cite{garrahan2010qu}.

\section{Effective interactions without time-reversal symmetry}
\label{sec:noneq}

This section surveys some results, mostly from the mathematical physics literature, which provide variational methods
for determining $u_\CC$ in systems without time-reversal symmetry, so that (\ref{equ:var-H}) does not apply.  For
systems of practical interest, we are proposing that
these results could be useful for (i) analytic bounds on dynamical free energies (for example, proving the existence of phase transitions in non-equilibrium systems, following the analysis of the time-reversible case~\cite{garrahan07,garrahan09}); (ii) variational analyses of effective interactions, as used in~\cite{jack14}; (iii) improved numerical procedures, for example obtaining an approximation to the auxiliary dynamics in order to improve sampling within a computational scheme.
Our purpose here is to highlight these opportunities so we mostly quote relevant results, 
referring to the literature for more detailed analysis and derivations.

\subsection{Optimal control theory}
\label{sec:opt}

We first state a general variational formula for the free energy $\psi(s)$, which may be viewed as a generalisation of
(\ref{equ:var-H}) for systems lacking time-reversal symmetry.  The variation is over sets of transition rates, which should
be chosen to reproduce the auxiliary rates (\ref{equ:waux}) as closely as possible. 
For type-$B$ observables, 
\begin{equation}
\psi(s) = \lim_{\tobs\to\infty} \left[ \min_{\{W^{\rm var}\}}\frac{1}{\tobs}  \left\langle  sB + \sum_{{\rm jumps}\; \CC'\leftarrow\CC}
\frac{L(\CC'\leftarrow \CC)}{W^{\rm var}(\CC'\leftarrow\CC)}
\right\rangle_{\rm var} \right]
\label{equ:control}
\end{equation}
where the variational parameters are (non-negative) rates $W^{\rm var}(\CC'\leftarrow\CC)$, the average is over 
a dynamical evolution under those rates starting from some arbitrary initial state, and
\begin{equation}
L(\CC\leftarrow\CC') = W^{\rm var}(\CC\leftarrow\CC')\left[
\ln \frac{W^{\rm var}(\CC\leftarrow\CC')}{W(\CC\leftarrow\CC')}-1\right] + W(\CC\leftarrow\CC') .
\label{equ:L-cost}
\end{equation}
We note here an equivalent way of writing the objective function in (\ref{equ:control}) above. By averaging over the number of jumps in any small time interval after time $t$, starting from the current configuration $\CC(t)$, one finds
\begin{equation}
\psi(s) = \lim_{\tobs\to\infty} \left[ \min_{\{W^{\rm var}\}}\frac{1}{\tobs} \int_0^{\tobs} \!\mathrm{d}t \left\langle s\,b(\CC(t)) + \sum_{\CC'}
L(\CC'\leftarrow \CC(t))
\right\rangle_{\rm var} \right]
\label{equ:control2}
\end{equation}
where we have also written out $B$ explicitly as a time integral.
The minima in (\ref{equ:control},\ref{equ:control2}) are obtained when the rates $W^{\rm var}$ are equal to the auxiliary rates defined
by (\ref{equ:waux}).  A derivation of this result will be sketched in Sec.~\ref{sec:2p5} below. We first give a brief discussion of its interpretation
and potential usefulness.

The variational principle (\ref{equ:control}) arises
 in ``optimal control theory''~\cite{fleming92,fleming05,hartmann12,kappen13}: the idea is that $W^{\rm var}$ is a ``controlled dynamics'' that 
should be optimised in order to realise the rare event of interest. 
The content of (\ref{equ:control}) is that the controlled process should minimise $s\langle B \rangle_{\rm var}$, while deforming the original
rates as little as possible.  
[Note that $L(\CC'\leftarrow\CC)$ resembles a relative entropy between the sets of transition rates, with $L=0$ if 
$W^{\rm var}(\CC'\leftarrow\CC)=W(\CC'\leftarrow\CC)$. 
In fact the final term in (\ref{equ:control2}) is exactly the small-$\dt$ limit of the relative entropy between the distributions of configurations reached from $\CC$ in a small time interval $\dt$ when the rates $W$ and $W^{\rm var}$ respectively are in force. For the rates $W$, this distribution is $P_{\dt}(\CC')=\dt W(\CC'\leftarrow \CC)$ for $\CC'\neq \CC$ and $P_{\dt}(\CC)=1-\dt r(\CC)$ otherwise; the relevant expressions for the rates $W^{\rm var}$ are analogous.] 
Since the maximum in (\ref{equ:control}) is obtained when $W^{\rm var}=W^{\rm aux}$, we may restrict the maximisation to rates
$W^{\rm var}(\CC'\leftarrow\CC) = W(\CC'\leftarrow\CC) \ee^{[\Delta V^{\rm var}(\CC) - \Delta V^{\rm var}(\CC')]/2} $ of the same form as $W^{\rm aux}$. Then $\Delta V^{\rm var}$
has the interpretation of an effective potential that pushes the system towards the rare event of interest.  In this context, (\ref{equ:control})
can be interpreted as an optimisation over the ``controlling field''  $\Delta V^{\rm var}$. 

In the case of diffusive processes, (\ref{equ:control}) has a particularly simple form: 
consider a model defined by a Langevin equation (or stochastic differential equation)
\begin{equation}
\dot{x} = K(x) + \eta
\end{equation}
where $K=K(x)$ is a force and $\eta$ is a white noise.  We then define a ``controlled process'' $\dot{x} = K - \partial_x V^{\rm var} + \eta$ where $V^{\rm var}=V^{\rm var}(x)$ is the controlling
potential. 
The idea is to discretize in time using a small time interval $\dt$.  For $x'\approx x$, one has
\begin{equation}
\frac{ L(x'\leftarrow x) }{ W^{\rm var}(x'\leftarrow x) } \approx (x'-x)(-\partial_x V)+\exp((x'-x)\partial_x V)-1
\end{equation}
Then averaging over $x'$ with weight $W^{\rm var}$ reduces this to $\dt (\partial_x V)^2/2 + O(\dt^2)$. 
Hence
\begin{equation}
\psi(s) =  \lim_{\tobs\to\infty} \min_{V^{\rm var}}\frac{1}{\tobs}  \int_0^{\tobs} \!\mathrm{d}t\, \left\langle   s\,b(x(t)) + \frac12 [\partial_x V(x(t))]^2 \right\rangle_{\rm var}
\end{equation}
where one seeks to simultaneously minimise the average of $sB$ and the magnitude of the controlling force $\partial_x V^{\rm var}$.
The relationships between optimal control and large deviations for diffusive systems
have been discussed in the physics literature~\cite{hartmann12,kappen13}, but while the results (\ref{equ:control},\ref{equ:control2}) for Markov chains
are known in the mathematical literature~\cite{dupuis}, they have not, to our knowledge, been applied very far in physics.

In terms of future applications,
it is clear that (\ref{equ:control}) gives bounds on $\psi$ and allows variational estimates of $W^{\rm aux}$.  
In principle this enables variational analyses of large deviations in non-equilibrium settings, similar
 to those described for time-reversible systems in Section~\ref{sec:kcm}.  However, there is an additional
difficulty associated with (\ref{equ:control}), which arises from the estimation of the average with respect to the variational (controlled) dynamics.
In the absence of detailed balance, these averages will typically need to be obtained by direct numerical simulation,
in which case convergence to the limit of large $\tobs$ may be non-trivial.

In the case of time-reversal symmetric ensembles, one can restrict to $W^{\rm var}$ that obey detailed balance, and  
(\ref{equ:control}) reduces to (\ref{equ:var-H}). 
To see this, replace the expectation value in (\ref{equ:control2}) by an average with respect to the steady state of the controlled dynamics $\mu^{\rm var}(\CC)\propto \ee^{-E(\CC)-\Delta V(\CC)}$.  The key point is that the logarithmic term in $L(\CC'\leftarrow\CC)$ yields $\sum_{\CC,\CC'}W^{\rm var}(\CC'\leftarrow\CC)\mu^{\rm var}(\CC)[\Delta V(\CC)-\Delta V(\CC')]/2$; using the detailed balance relation $W^{\rm var}(\CC'\leftarrow\CC)\mu^{\rm var}(\CC)=W^{\rm var}(\CC\leftarrow\CC')\mu^{\rm var}(\CC')$ and interchanging the summation variables shows that this term vanishes. Finally using $r(\CC) = \sum_{\CC'} W(\CC'\leftarrow \CC)$, Eq.~(\ref{equ:control2}) reduces to
\begin{equation}
\psi = \min_{\{W^{\rm var}\}}
\sum_{\CC} \left[ sb(\CC) + r(\CC) - \sum_{\CC'} W^{\rm var}(\CC'\leftarrow\CC)\right]
\mu^{\rm var}(\CC)
\end{equation}
which is the same as (\ref{equ:var-H}). 

We highlight two other potential routes for application of (\ref{equ:control}).  First, it can provide simple bounds on $\psi$ by appropriate
simple choices of $W^{\rm var}$.  For example if one biases by the total
activity (number of spin flips), and the system has a configuration with sub-extensive escape rate [there exists a sequence of configurations $\CC_N$ in systems of increasing size $N$ such that $r(\CC_N)/N\to0$ as $N\to\infty$], 
then $\lim_{N\to\infty}\psi(s)/N\leq0$ and hence (given weak conditions on properties of the steady state)
 there must be a dynamical phase transition at $s=0$.  This is a non-equilibrium analogue of results proven
for kinetically constrained models of the glass transition~\cite{garrahan07,garrahan09}.  It is relevant for
exclusion processes, where the same result may be derived either by exact solution~\cite{bodineau05} or within fluctuating hydrodynamics~\cite{bertini14}.  But the method
based on (\ref{equ:control}) is both very simple and very general.
Second, there should be possibilities of using (\ref{equ:control}) in numerical schemes, for example by generalising
the method of Nemoto and Sasa~\cite{nemoto14}.  This possibility remains to be explored.

Finally one could also consider finite-$\tobs$ analogues of (\ref{equ:control2}). We define $\phi(\CC,\tobs)$ as the minimum value of the objective function on the r.h.s.\ of (\ref{equ:control2}) when starting from a given configuration $\CC$. It is then not difficult to argue that $\phi(\CC,\tobs)=\psi(s) + \Delta V_\CC/(2\tobs)$ for large $\tobs$, up to corrections that decay exponentially with $\tobs$. If one allows the variational rates $W^{\rm var}$ to depend on time then one can also obtain a closed form for the evolution equation of the $\phi(\CC,\tobs)$. Thus it may be possible to obtain the effective interactions from the finite-$\tobs$ behaviour of the optimal cost $\phi(\CC,\tobs)$ in the control theory approach.

\subsection{Large deviations at ``level-2.5''}
\label{sec:2p5}

To understand the origin of the variational result (\ref{equ:control}), it is useful to consider the large deviations of a very general set of observables~\cite{maes08}.  
For a given trajectory $\CC(t)$, we define the \emph{empirical current} which is a set of numbers $Q(\CC'\leftarrow \CC)$, 
obtained by counting the jumps (transitions) that between each pair of configurations, and dividing by $\tobs$.  Similarly, the
\emph{empirical measure} is a set of numbers $\mu(\CC)$ given by the fraction of time
that the trajectory spent in each configuration. Note that when we consider the number of jumps into and out of each configuration $\CC$ along any trajectory, these two numbers must be equal or, if $\CC$ is the initial or final configuration, differ by at most unity. Due to the division by $\tobs$ in the definition of $Q(\CC'\leftarrow \CC)$ this difference vanishes for large $\tobs$, leading to the balance condition
\begin{equation}
\sum_{\CC'} Q(\CC'\leftarrow \CC) = \sum_{\CC} Q(\CC\leftarrow\CC').
\label{equ:qbal}
\end{equation}

\subsubsection{Main result}

The observables $\mu$ and $Q$ are very high-dimensional objects if the state space is large, but as long as there are a finite number of them they obey a large deviation
principle whose explicit rate function is known, for both biased and unbiased ensembles of trajectories. 
We first state the result~\cite{chetrite14-l2p5}:
for sufficiently large $\tobs$, 
one has
$p(\mu,Q) \sim \ee^{-\tobs I(\mu,Q)}$ with 
\begin{equation}
I(\mu,Q) =  
\sum_{\CC,\CC'} \left\{
Q(\CC'\leftarrow \CC) \left[ \log \frac{ Q(\CC'\leftarrow \CC) }{ W(\CC'\leftarrow \CC) \mu(\CC) } - 1\right] + W(\CC'\leftarrow \CC) \mu(\CC) 
\right\}
\label{equ:2p5}
\end{equation}
In an ensemble biased by a type-$B$ observable according to (\ref{equ:s-ens}), all trajectories with a given $\mu$ and $Q$ are reweighted by the same factor $\ee^{-sB}=\exp(-s \sum_\CC b(\CC)\mu(\CC))$, so after including the normalization factor $1/\ee^{-\tobs \psi(s)}$ one has $p(\mu,Q) \sim \ee^{-\tobs [I(\mu,Q,s)-\psi(s)]}$
with 
\begin{equation}
I(\mu,Q,s) = I(\mu,Q) + s \sum_{\CC} b(\CC) \mu(\CC)
\label{equ:2p5s}
\end{equation}

The result (\ref{equ:2p5}) is known as a ``level-2.5'' large deviation principle (LDP) since it is intermediate between an LDP for the empirical measure 
(known as level 2) and a full LDP for trajectories (known as level 3).
A review of large deviations at level-2.5 is given in~\cite{chetrite14-l2p5}, while rigorous analysis of the case with countably infinite state
spaces is given in~\cite{bertini12-flow,bertini12-dv}, including a proof of (\ref{equ:2p5}).

\subsubsection{Connection to the auxiliary process}

The typical empirical measure and current within the biased ensemble of (\ref{equ:s-ens})
can be obtained by minimisation  over $\mu$ and $Q$ of the rate function $I(\mu,Q,s)$ in (\ref{equ:2p5}).  
The key point for our purposes is that this allows a variational determination of the auxiliary model of (\ref{equ:waux}).
We first cast the minimisation over $\mu,Q$ as a minimisation over a ``variational auxiliary model'':
a model for which typical trajectories have current $Q$ and measure $\mu$.  The transition rates of this model then have to be
\begin{equation}
W^{\rm var}(\CC'\leftarrow \CC) = Q(\CC'\leftarrow \CC) / \mu(\CC)  .
\end{equation}
As one would expect, the balance constraint (\ref{equ:qbal}) on $Q$ then ensures that the steady state of the process defined by the rates $W^{\rm var}$ is $\mu$.

We rewrite (\ref{equ:2p5}) as
\begin{equation}
I(\mu,Q) =  \sum_{\CC,\CC'}\left\{ W^{\rm var}(\CC'\leftarrow \CC)
\left[ \log \frac{ W^{\rm var}(\CC'\leftarrow \CC) }{ W(\CC'\leftarrow \CC) } -1\right] + W(\CC'\leftarrow \CC) \right\}  \mu(\CC) .
\label{equ:2p5-waux}
\end{equation}
Including the bias term as in (\ref{equ:2p5s}) gives the function $I(\mu,Q,s)$, which is minimised by the $(\mu,Q)$ that are most likely within the biased ensemble: we denote their values by $(\mu^*,Q^*)$.
The variational rates $W^{\rm var}$ at this minimum of $I(\mu,Q,s)$ define a model for which
typical trajectories have $(\mu,Q)=(\mu^*,Q^*)$ so they must be
exactly the auxiliary rates $W^{\rm aux}$ associated with the biased ensemble.  

Moreover, the minimal value of $I(\mu,Q,s)$ itself is just the dynamical free energy: $\psi(s)=I(\mu^*,Q^*,s)$. 
This follows from a contraction principle~\cite{touchette09} because 
the observable $B=\tobs\sum_\CC b(\CC)\mu(\CC)$ is a simple function of the empirical measure: recall from (\ref{equ:z-psi}) 
that $\langle \ee^{-sB} \rangle_0 \sim \ee^{-\tobs\psi(s)}$.  Hence, decomposing the average into contributions from all possible $\mu,Q$,
one has
\begin{equation}
\langle \ee^{-sB} \rangle \sim \max_{\mu,Q} \left[ \ee^{-\tobs I(\mu,Q)} \ee^{-s\tobs\sum_\CC b(\CC)\mu(\CC)}\right] \sim \max_{\mu,Q} \ee^{-\tobs I(\mu,Q,s)}.
\end{equation}

Summarising the ingredients so far, we have $\psi(s)=\min_{\mu,Q} I(\mu,Q,s)$ where the minimization can equivalently be done over the variational rates $W^{\rm var}$ rather than $\mu$ and $Q$. The final step in the argument is to realize that in the large $\tobs$-limit, the average over $\CC(t)$ in the optimal control formulation (\ref{equ:control2}) above becomes an average over the stationary measure $\mu(C)$, making the penalty term $L$ equal to $I(\mu,Q)$ as rewritten in (\ref{equ:2p5-waux}). Thus
 the variational principle (\ref{equ:control}) follows from the general level-2.5 result~(\ref{equ:2p5}).

\subsubsection{Derivation of (\ref{equ:2p5-waux})} 

A derivation of (\ref{equ:2p5-waux}) for the case $s=0$ is given in~\cite{maes08}.  
Here we outline their argument.   The empirical current and measure $(\mu,Q)$ are typical for the process $W^{\rm var}$, 
but they are not typical for the original process $W$. For a given trajectory, one may obtain an expression for the ratio
$P[\CC(t);0]/P[\CC(t);{\rm var}]$, usng representations analogous to (\ref{equ:ptraj}).  Then one writes the probability of
observing an empirical current and measure in the unbiased process as 
\begin{equation}
\ee^{-\tobs I(\mu,Q)} \sim \sum_{\CC(t) | \mu,Q} P[\CC(t);0] = \sum_{\CC(t) | \mu,Q} \frac{P[\CC(t);0]}{P[\CC(t);{\rm var}]}\,P[\CC(t);{\rm var}]
 \sim \left\langle \frac{P[\CC(t);0]}{P[\CC(t);{\rm var}]} \right\rangle_{\rm var}
\end{equation}
The summations in this equation should be interpreted as path integrals over all trajectories that are compatible with an empirical current and measure $(\mu,Q)$.
The average on the r.h.s.\ is with respect to the $W^{\rm var}$ process: the restriction to a given $(\mu,Q)$ can
be omitted here since this average is already dominated by such trajectories, which are typical for that process.
Using the explicit form of the ratio to be averaged then yields (\ref{equ:2p5}).  The analysis of~\cite{maes08} considered only the case $s=0$, but the
general result of (\ref{equ:2p5s}) follows immediately as explained above, because the effect of the bias in (\ref{equ:s-ens}) can be re-written as a bias that
depends only on the empirical measure.
  
\subsection{Large deviations at level-2}
\label{sec:level2}

Finally, we recall a classical result of Donsker and Varadhan~\cite{dv75} for large
deviations of the empirical measure. Without bias, these  
satisfy $p(\mu) \sim \ee^{-\tobs J(\mu)}$ with \begin{equation}
J(\mu) = \max_{\rho} \left\{ - \sum_{\CC,\CC'} \rho(\CC') W(\CC'\leftarrow\CC) \frac{\mu(\CC)}{\rho(\CC)}  + \sum_{\CC} r(\CC) \mu(\CC).
\right\}
\label{equ:DV}
\end{equation}
where the maximisation is over a set of variational parameters $\rho(\CC)>0$. In the $B$-biased ensemble the relevant large deviation function just needs to add the effect of the bias as before, giving $p(\mu) \sim \ee^{-\tobs [J(\mu,s)-\psi(s)]}$ with 
\begin{equation}
J(\mu,s) = J(\mu) + \sum_{\CC} s\,b(\CC)\mu(\CC).
\label{equ:DV-bs}
\end{equation}
The corresponding expression for ensembles biased by type-$A$ observables is given in~\cite[Appendix C]{garrahan09}.
Subsequent minimisation over $\mu$ yields the dynamical
free energy $\psi$, and the $\rho(\CC)$ at the minimum are the $u(\CC)$ associated with the auxiliary dynamics of  Eq.~(\ref{equ:waux}). 

The result (\ref{equ:DV}) can 
be obtained by minimisation of (\ref{equ:2p5}) over $Q$, subject to the balance constraints (\ref{equ:qbal}). Calling the minimum value $J(\mu)$, we want to show that it can be obtained alternatively from the maximisation problem (\ref{equ:DV}). This can be done using Lagrangian duality: the Lagrangian for the original minimisation is
\begin{equation}
L(\mu,Q,\lambda)=I(\mu,Q)+\sum_{\CC,\CC'} \lambda(\CC) [Q(\CC'\leftarrow \CC) - Q(\CC \leftarrow \CC')] .
\end{equation}
The dual Lagrangian is then defined as $\tilde{L}(\mu,\lambda) = \min_{Q} L(\mu,Q,\lambda)$. Since for any $Q$ satisfying (\ref{equ:qbal}) one has $L(\mu,Q,\lambda)=I(\mu,Q)$, it follows that $\tilde{L}(\mu,\lambda)
\leq
J(\mu)$. Since the equality holds for the optimal $Q$, one has the dual representation $J(\mu)=\max_\lambda \tilde L(\mu,\lambda)$. Now setting the derivative of $L(\mu,Q,\lambda)$ to zero to find $\tilde L(\mu,\lambda)$ gives
\begin{equation}
\log \frac{Q(\CC'\leftarrow \CC)}
{W(\CC'\leftarrow \CC) \mu(\CC)} = \lambda(\CC')-\lambda(\CC)
\end{equation}
Substituting back into $L(\mu,Q,\lambda)$, the log term cancels with the Lagrange multiplier contribution and one is left with
\begin{equation}
\tilde{L}(\mu,\lambda)
\sum_{\CC,\CC'} \left\{
-W(\CC'\leftarrow \CC)\mu(\CC)\ee^{\lambda(\CC)-\lambda(\CC')} + W(\CC'\leftarrow \CC) \mu(\CC) 
\right\}
\end{equation}
Identifying $\rho(\CC)=\ee^{-\lambda(\CC)}$ and carrying out the sum over $\CC'$ in the second term then gives (\ref{equ:DV}) as desired.

To our knowledge, Eq.~(\ref{equ:DV-bs}) has had limited application for estimation of $\psi(s)$ and the $u(\CC)$.  One obstacle is
that this requires a \emph{maximisation} over $\rho$, followed by a minimisation over $\mu$.  For this reason, straightforward bounds on $\psi$ 
are not directly available, unlike the case of (\ref{equ:2p5}) where one minimises over both $\mu$ and $Q$.

\section{Outlook}
\label{sec:conc}

We have summarised a range of analytical and numerical results related to the effective potentials encoded by (\ref{equ:eff-pot}).  
Section~\ref{sec:illustrate} reviews some previous
results where these effective potentials have been estimated, mostly in time-reversal symmetric ensembles.  Section~\ref{sec:noneq} shows
how the effective potentials can be interpreted in terms of the controlling forces that achieve rare events most 
efficiently, in the sense of the ``objective function'' $L$ in (\ref{equ:L-cost}).  
We have discussed how the variational results described in Sections~\ref{sec:opt}
and~\ref{sec:2p5} might be useful for generalising these kinds of method to systems without detailed balance, and for developing
new numerical methods, possibly following Ref.~\cite{nemoto14}.  The application of these results to biased ensembles for
open quantum systems~\cite{garrahan2010qu} might also provide useful insights.

Another general challenge coming from biased ensembles is the description of biased states that are inhomogeneous in space and time.  The ``addivity
principle'' leads to some exact results in homogeneous systems, but an accurate description of spatially inhomogeneous (phase-separated)
states remains outstanding in some cases.  Biased ensembles also support ``travelling-wave'' states which are inhomogeneous in both space 
and time~\cite{bodineau05,bertini06}: it might be useful to investigate variational techniques based on (\ref{equ:2p5}) in order to address these problems.

From a fundamental point of view, the relation between effective interactions and the thermodynamic limit 
is also important. Biased ensembles in general will be characterised by some stationary measure $\mu(\CC)$. Restricting for convenience to systems with time-reversal symmetry
one then expects that this has the form $\mu(\CC) \propto \mu_0(\CC) \ee^{-\Delta V_\CC}$, where $\mu_0$ is the stationary distribution
of the unbiased process, and $\Delta V_{\CC}$ an effective potential.  However, in the thermoydynamic limit,
a question arises as to whether the measure $\mu$ is ``Gibbsian''~\cite{gibbsian93,maes99}:
that is, whether $\Delta V$ can be written as a well-defined sum of interaction terms of increasing range.  If such a description is not possible, 
even the definition of effective interactions becomes problematic in the thermodynamic limit.  If one considers large deviations
of the total energy (type-$B$) in the Ising model, there is evidence that the resulting effective interactions may not be Gibbsian~\cite{js10}.
It is also not clear how whether the limits of large system size and large-$\tobs$ should commute in such cases, and what consequences this might have.
It would be interesting to analyse these questions further in future work.

\section*{Acknowledgments}

We thank Raphael Ch\'{e}trite, Hugo Touchette, Carsten Hartmann, Vivien Lecomte, Fred van Wijland, Juan Garrahan, and David Chandler for many useful
discussions on the issues discussed here.  RLJ thanks the EPSRC for support through grant EP/I003797/1.


\begin{thebibliography}{}


\bibitem{auer01}
S.~Auer and D.~Frenkel, Nature {\bf 409}, 6823 (2001).
\bibitem{sear07}
R.~P.~Sear, J. Phys.: Cond. Matt. {\bf 19}, 033101(2007).
\bibitem{ren05}
W.~Ren, E.~vanden-Eijnden, P.~Maragakis and W. E, J. Chem. Phys. {\bf 123}, 134109 (2005).

\bibitem{hanggi90}
P.~Hanggi, P.~Talkner and M.~Borkovec, Rev. Mod. Phys. {\bf 62}, 241 (1990).
\bibitem{wales-book}
D. J. Wales, {\it Energy Landscapes} (Cambridge University Press, Cambridge, 2003).
\bibitem{e05}
W. E, W. Ren and E. vanden-Eijden, J. Phys. Chem. B {\bf 109}, 6688 (2005).

\bibitem{tps}
P.~G.~Bolhuis, D.~Chandler, C.~Dellago and P.~L.~Geissler,
Ann. Rev. Phys. Chem. {\bf53}, 291 (2002).
\bibitem{allen06}
R.~J.~Allen, D.~Frenkel and P.~R.~ten Wolde, J. Chem. Phys. {\bf 124}, 024102 (2006).

\bibitem{touchette09}
H. Touchette, Phys. Rep. {\bf 478}, 1 (2009).

\bibitem{galla95}
G.Gallavotti and E.~G.~D.~Cohen, Phys. Rev. Lett. {\bf 74}, 2694 (1995).
\bibitem{lebowitz99}
J.~L.~Lebowitz and H.~Spohn, J. Stat. Phys. {\bf 95}, 333 (1999).

\bibitem{garrahan07}
J.~P.~Garrahan, R.~L.~Jack, V.~Lecomte, E.~Pitard, K.~van~Duijvendijk and F.~van~Wijland,
Phys. Rev. Lett. {\bf 98}, 195702 (2007)
\bibitem{garrahan09}
J.~P.~Garrahan, R.~L.~Jack, V.~Lecomte, E.~Pitard, K.~van Duijvendijk, and F.~van Wijland,
J. Phys. A {\bf 42}, 075007 (2009).
\bibitem{hedges09}
L.~O.~Hedges, R.~L.~Jack, J.~P.~Garrahan and D.~Chandler, 
Science {\bf 323}, 1309 (2009).
\bibitem{speck12}
T. Speck and D. Chandler, J. Chem. Phys. {\bf 136}, 184509 (2012)

\bibitem{weber13}
J.~K.~Weber, R.~L.~Jack and V.~S.~Pande, J. Am. Chem. Soc. {\bf 135}, 5501 (2013)
\bibitem{mey14}
A.~S.~J.~S.~Mey, P.~L.~Geissler and J.~P.~Garrahan, Phys. Rev. E {\bf 89}, 032109 (2014)
\bibitem{weber14}
J.~K.~Weber, R.~L.~Jack, C.~R.~Schwandtes and V.~S.~Pande,
Biophys. J {\bf 107}, 974 (2014)

\bibitem{tailleur07}
J.~Tailleur and J.~Kurchan, Nat. Physics {\bf 3}, 203 (2007)
\bibitem{lam09}
K.-D. N. T. Lam, J. Kurchan and D.~Levine, J. Stat. Phys. {\bf 137}, 1079 (2009)

\bibitem{derrida98}
B.~Derrida and J.~L.~Lebowitz, Phys. Rev. Lett. {\bf 80}, 209 (1998)
\bibitem{bertini01}
L. Bertini, A. De Sole, D. Gabrielli, G. Jona-Lasinio, and C. Landim,
Phys. Rev. Lett. {\bf 87}, 040601 (2001)
\bibitem{bodineau04}
T.~Bodineau and B.~Derrida, Phys. Rev. Lett. {\bf 92}, 180601 (2004).
\bibitem{bertini14}
L. Bertini, A.~De Sole, D.~Gabrielli, G.~Jona-Lasinio and C.~Landim,
arXiv:1404.6466 (2014)


\bibitem{dv75}
M.~D.~Donsker and S.~R.~S.~Varadhan, Commun. Pure Appl. Math. {\bf 28}, 1 (1975)

\bibitem{evans04}
R.~M.~L.~Evans, Phys. Rev. Lett. {\bf 92}, 150601 (2004).
\bibitem{evans05}
R.~M.~L.~Evans, J. Phys. A {\bf 38}, 293 (2005).

\bibitem{maes08}
C.~Maes and K.~Netocny, EPL {\bf 82}, 30003 (2008)

\bibitem{js10}
R.~L.~Jack and P.~Sollich, Prog.~Theor.~Phys.~Supp.~{\bf 184}, 304 (2010)

\bibitem{chetrite14-hp}
R. Ch\'{e}trite and H. Touchette, Ann. Henri Poincar\'{e}, in press (2014). {\tt doi:10.1007/s00023-014-0375-8}.

\bibitem{fleming92}
W.~H.~Fleming, \emph{Stochastic control and large deviations}, 
in \emph{Future Tendencies in Computer Science, Control and Applied Mathematics}, pages 291-300
(Springer, Berlin, 1992).  {\tt doi:10.1007/3-540-56320-2\char`_66}.

\bibitem{fleming05}
W.~H.~Fleming and H. M. Soner, \emph{Controlled Markov Processes and Viscosity Solutions}
(Springer, Berlin, 2005)
\bibitem{hartmann12}
C.~Hartmann and C.~Sch\"{u}tte, J. Stat. Mech. (2012), P11004
\bibitem{kappen13}
V.~Y.~Chernyak, M.~Chertkov, J.~Bierkens and H.~J.~Kappen, 
J. Phys. A {\bf 47}, 022001 (2013)

\bibitem{chetrite14-l2p5}
A.~C.~Barato and R.~Chetrite, arXiv:1408.5033.

\bibitem{lecomte07jsp}
V.~Lecomte, C.~Appert-Roland and F. van Wijland, J. Stat. Phys. 127, 51 (2007)

\bibitem{chetrite13}
R.~Ch\'{e}trite and H.~Touchette, Phys. Rev. Lett. {\bf 111}, 120601 (2013)

\bibitem{stroock}
D.~W.~Stroock, {\it An Introduction to Markov Processes} (Springer, Berlin/Heidelberg, 2005)

\bibitem{cameron14}
M.~Cameron and E. Vanden-Eijnden, J. Stat. Phys. {\bf 156}, 427 (2014).

\bibitem{rs03}
F.~Ritort and P.~Sollich, Adv. Phys. {\bf 52}, 219 (2003)
\bibitem{gst11}
J.~P.~Garrahan, P.~Sollich and C.~Toninelli, {\it Kinetically constrained models}, Ch. 10
in {\it Dynamical heterogeneities in glasses, colloids, and granular media}, Eds.: L. Berthier, G. Biroli, J-P Bouchaud, L. Cipelletti and W. van Saarloos (Oxford University Press, Oxford 2011).
\bibitem{gc10}
D. Chandler and J.~P.~Garrahan, Ann. Rev. Phys. Chem. {\bf 61}, 191 (2010).

\bibitem{jack10}
R.~L.~Jack and J.~P.~Garrahan, Phys. Rev. E {\bf 81}, 011111 (2010)

\bibitem{jack14}
R.~L.~Jack and P.~Sollich, J. Phys. A {\bf 47}, 015003 (2014)

\bibitem{popkov11}
V.~Popkov and G.~M.~Sch\"utz, J. Stat. Phys. {\bf 142}, 627 (2011)
\bibitem{jack14-hyper}
R.~L.~Jack, I.~R.~Thompson and P. Sollich, arXiv:1409.3986. 
\bibitem{torquato03}
S.~Torquato and F.~H.~Stillinger,
Phys. Rev, E {\bf 68}, 041113 (2003).
\bibitem{gabrielli2003}
A.~Gabrielli, B.~Jancovici, M.~Joyce, J.~L.~Lebowitz, L.~Pietronero and F. Sylos Labini,
Phys. Rev. D {\bf 67}, 043506 (2003).
\bibitem{zachary}
C.~E.~Zachary, Y.~Jiao and S.~Torquato, Phys. Rev. Lett. {\bf 106}, 178001 (2011)
\bibitem{chicken2014}
Y.~Jiao, T.~Lau, H.~Hatzikirou, M.~Meyer-Hermann, J.~C.~Corbo and S.~Torquato,
Phys. Rev. E {\bf89}, 022721 (2014)

\bibitem{bertini05}
L. Bertini, A. De Sole, D. Gabrielli, G. Jona-Lasinio, and C. Landim,
Phys. Rev. Lett. {\bf 94}, 030601 (2005)
\bibitem{bodineau05}
T.~Bodineau and B.~Derrida,
Phys. Rev. E {\bf 72}, 066110 (2005)

\bibitem{hurtado14}
P.~I.~Hurtado, C.~P.~Espigares, J.~J.~del Pozo, P.~L.~Garrido,
J. Stat. Phys. {\bf 154}, 214 (2014)

\bibitem{merolle05}
M.~Merolle, J.~P.~Garrahan and D.~Chandler, PNAS {\bf 102}, 10837 (2005).
\bibitem{crooks01}
G.~E.~Crooks and D.~Chandler, Phys. Rev. E {\bf 64}, 026109 (2001)

\bibitem{giardina06}
C.~Giardina, J.~Kurchan and L.~Peliti, Phys. Rev. Lett. {\bf 96}, 120603 (2006)
\bibitem{lecomte07jsm}
V.~Lecomte and J.~Tailleur, J. Stat. Mech. {\bf (2007)}, P03004

\bibitem{nemoto14}
T.~Nemoto and S.~Sasa, Phys. Rev. Lett. {\bf 122}, 090602 (2014)

\bibitem{garrahan2010qu}
J.~P.~Garrahan and I.~Lesanovsky, Phys. Rev. Lett, {\bf 104}, 160601 (2010)

\bibitem{dupuis}
P.~Dupuis and R.~S.~Ellis, \emph{A Weak Convergence Approach to the Theory of Large Deviations}
(Wiley, New York, 1997).

\bibitem{bertini12-flow}
L. Bertini, D. Gabrielli and A. Faggionnato, arXiv:1210.2004 (2012).

\bibitem{bertini12-dv}
L. Bertini, A. Faggionato, and D. Gabrielli. 
arXiv:1212.6908 (2012).

\bibitem{bertini06}
L.~Bertini, A.~De Sole, D.~Gabrielli, G.~Jona-Lasinio
and C.~Landim, J. Stat. Phys. {\bf 123}, 237 (2006).

\bibitem{gibbsian93}
A.~C.~D. van Enter, R. Fern\'{a}ndez and A.~D.~Sokal, J. Stat. Phys. {\bf 72}, 879 (1993)
\bibitem{maes99}
C.~Maes, F.~Redig and A.~van Moffaert, J. Stat. Phys.~{\bf 96}, 69 (1999)

\end{thebibliography}
\end{document}